%% file: main.tex
\documentclass[aps,prc,reprint,nofootinbib,superscriptaddress,showkeys,floatfix,longbibliography]{revtex4-2}

\usepackage[T1]{fontenc}
\usepackage[utf8]{inputenc}
\usepackage{lmodern}
\usepackage{amsmath,amssymb,bm}
\usepackage{booktabs}
\usepackage{graphicx}
\usepackage{xcolor}
\usepackage{microtype}
\usepackage{placeins}
\usepackage{hyperref}

\graphicspath{{figures/}}
\hypersetup{
  colorlinks=true,
  linkcolor=blue!55!black,
  citecolor=blue!55!black,
  urlcolor=blue!55!black,
  pdftitle={Signed Sound-Speed Deformations of BSk24-Anchored Barotropes: Neutron-Star Response},
  pdfauthor={Ioannis Papathanasiou},
  pdfsubject={Physics methods for controlled BSk24-anchored sound-speed deformations},
  pdfkeywords={neutron stars, equation of state, sound speed, thermodynamic consistency, tidal deformability, BSk24}
}

\input{tables/generated_values}

\begin{document}

\title{Signed Sound-Speed Deformations of BSk24-Anchored Barotropes: Neutron-Star Response}

\author{Ioannis Papathanasiou}
\email{ipapatha@auth.gr}
\homepage{https://orcid.org/0009-0003-2792-8781}
\affiliation{Department of Theoretical Physics, School of Physics,
Aristotle University of Thessaloniki, 54124 Thessaloniki, Greece}

\begin{abstract}
\begin{description}
\item[Background] Localized structure in the equilibrium squared sound speed
can affect neutron-star observables, but modifying \(dP/d\epsilon\) requires
integration into a thermodynamically consistent barotrope and assessment over
its complete declared domain.
\item[Purpose] We quantify the response to positive and negative localized
sound-speed changes while holding fixed the analytic representation of the
unified BSk24 baseline, deformation geometry, admission criteria,
reconstruction, and fixed-mass comparison protocol.  Only the dimensionless
amplitude \(A\) varies.
\item[Method] A Gaussian profile with quintic smootherstep activation is added
to \(c_{\mathrm{s}}^2\).  Raw proposals that violate \(P>0\) or
\(0<c_{\mathrm{s}}^2\leq1\) anywhere in the retained domain are rejected
before reconstruction.  The remaining cases are reconstructed as effective
cold one-fluid barotropes and assessed with thermodynamic-closure diagnostics.
Formal-interval members are evolved with the
Tolman--Oppenheimer--Volkoff and tidal equations and compared at fixed
gravitational mass using a common finite-domain surface cutoff.
\item[Results] The \(A=0\) thermodynamic fields and saved stellar sequences
reproduce the undeformed control exactly.  A localized sound-speed change
leaves a persistent pressure offset, relocates the fixed-mass central state,
and produces a mass-dependent, sign-asymmetric finite-amplitude response.  For
\(A=-0.8\) and \(+0.8\), respectively,
\(\Delta\Lambda/\Lambda_0=(-46.02\%,+35.33\%)\) at
\(1.4\,M_\odot\) and \((-63.52\%,+87.17\%)\) at
\(2.0\,M_\odot\).  At the selected reporting endpoints, the reached stiffened
feature remains core connected, whereas the softened feature is sampled as an
off-center shell at high mass.
\item[Conclusions] Even a localized change in \(dP/d\epsilon\) has nonlocal
barotropic and stellar consequences through pressure integration and
central-state relocation.  The results characterize phenomenological
BSk24-anchored effective barotropes, not predictions of the BSk24 functional,
a baseline-independent response law, or evidence for a microscopic
composition or phase transition.
\end{description}
\end{abstract}

\maketitle

\input{sections/01_introduction}
\input{sections/02_framework}
\input{sections/03_experiment}
\input{sections/04_results}
\FloatBarrier
\input{sections/05_scope_conclusions}
\input{sections/appendices}

\FloatBarrier
\bibliography{references}

\end{document}

%% file: tables/generated_values.tex
\newcommand{\RawAmplitudeCount}{110}
\newcommand{\RawAcceptedCount}{106}
\newcommand{\RawRejectedCount}{4}
\newcommand{\StellarAmplitudeCount}{105}
\newcommand{\NegativeStellarCount}{49}
\newcommand{\PositiveStellarCount}{57}

\newcommand{\FormalAmplitudeLower}{-1.9733}
\newcommand{\FormalAmplitudeUpper}{4.3745}
\newcommand{\FormalAmplitudeLowerEnergy}{309.52}
\newcommand{\FormalAmplitudeUpperEnergy}{414.86}
\newcommand{\SelectedNegativeAmplitude}{-1.6}
\newcommand{\SelectedPositiveAmplitude}{2}
\newcommand{\MatchedAmplitude}{0.8}

\newcommand{\RealizedPeakEnergy}{383.8}
\newcommand{\RealizedCentroid}{412.0}
\newcommand{\RealizedFWHM}{296.3}

\newcommand{\NMatch}{0.08437}
\newcommand{\PMatch}{0.30245}
\newcommand{\MuMatch}{951.83}
\newcommand{\BSkTwentyFourNSat}{0.1578}

\newcommand{\RealizedPeakDensityOverNSat}{2.435}

\newcommand{\RealizedFWHMLowerDensityOverNSat}{1.629}
\newcommand{\RealizedFWHMUpperDensityOverNSat}{3.291}

\newcommand{\PositivePeakDeltaCsTwo}{0.2959}
\newcommand{\NegativePeakDeltaCsTwo}{-0.2367}
\newcommand{\PositivePressureMemory}{91.96}
\newcommand{\NegativePressureMemory}{-73.56}
\newcommand{\PositiveLambdaPercentOneFour}{72.1}

\newcommand{\NegativeLambdaPercentOneFour}{-76.4}

\newcommand{\PositiveLambdaPercentTwoZero}{189.7}

\newcommand{\NegativeLambdaPercentTwoZero}{-89.5}

\newcommand{\EpsCNegativeTwoZero}{1505}

\newcommand{\PositiveSupportOuterRadiusTwoZero}{0.713}
\newcommand{\PositiveSupportOuterMassTwoZero}{0.543}
\newcommand{\NegativeSupportInnerRadiusTwoZero}{0.843}
\newcommand{\NegativeSupportOuterRadiusTwoZero}{0.935}
\newcommand{\NegativeSupportInnerMassTwoZero}{0.818}
\newcommand{\NegativeSupportOuterMassTwoZero}{0.971}
\newcommand{\ReferenceResolvedMaxima}{62}
\newcommand{\ReferenceDomainLimited}{34}
\newcommand{\ReferenceAmbiguousMaxima}{7}
\newcommand{\ReferenceFailedMaxima}{2}
\newcommand{\TidalBookkeepingResidual}{1.47\times10^{-9}}

%% file: sections/01_introduction.tex
\section{Introduction}
\label{sec:introduction}

The structure of a neutron star is controlled by how pressure \(P\) rises
with total energy density \(\epsilon\).  For a cold barotrope, the corresponding
local stiffness is measured by the equilibrium squared sound speed
\begin{equation}
  c_{\mathrm{s}}^{2}(\epsilon)=\frac{dP}{d\epsilon},
  \label{eq:sound-speed-definition}
\end{equation}
which is dimensionless when \(c=1\)
\cite{ComposeManual2022,Tews2018}.  Sound-speed representations complement
piecewise-polytropic, spectral, and continuous-fit parameterizations of dense
matter
\cite{Read2009,Lindblom2010Spectral,Lindblom2018Causal,Tews2018,Greif2019,
Servignat2024}.  An analytic baseline is particularly useful for a controlled
comparison because its derivatives and retained domain can remain explicit
\cite{HaenselPotekhin2004}.

Localized sound-speed structure has been studied through density-dependent
extensions and Gaussian components in equation-of-state inference and
parameterization comparisons \cite{Tews2018,Greif2019}; through bumps, spikes,
steps, plateaus, and kinks in stellar and tidal studies
\cite{Tan2022Extreme,Tan2022BinaryLove}; and through sharp structures and
ensembles motivated by phase-transition searches
\cite{Somasundaram2023,Mroczek2023}.  Constrained Gaussian-process bridges
provide a complementary ensemble construction with global stability,
causality, and thermodynamic-consistency restrictions
\cite{Gorda2026ConstrainedGP}.  These studies establish that structured sound
speeds can have observable consequences while also showing that a sound-speed
feature alone does not identify a microscopic constituent or phase transition.

The present study instead asks a narrower comparative question: how does one
signed feature propagate when the baseline and every other design choice are
held fixed?  We use the analytic representation of the unified BSk24 equation
of state because it supplies a single crust--core baseline with documented
coefficients, derivatives, domain limits, and published stellar comparisons
\cite{Goriely2013HFB24,Pearson2018,Pearson2019Erratum,PotekhinIoffe}.
Following exploratory comparisons, one ramped-Gaussian geometry is fixed for
the final campaign, and only the sign and magnitude of \(A\) change.  The
specific methodological contribution is the combination of a fixed-baseline
signed intervention, an analytic pressure primitive, fail-closed mechanical
and causal admission over the complete retained domain, an exact
zero-amplitude control, and an audit of the feature's radial support.

The calculation follows the chain
\begin{equation}
\begin{aligned}
\Delta c_{\mathrm{s}}^2
&\longrightarrow P_A(\epsilon)=P_0(\epsilon)+\Delta P
 \longrightarrow \{n_{\mathrm B},\mu_{\mathrm B}^{\rm eff}\}\\
&\longrightarrow \{R,\epsilon_c,k_2,\Lambda\}.
\end{aligned}
\label{eq:response-chain}
\end{equation}
Accepted cases are compared at fixed gravitational mass through both global
observables and the position of the realized feature inside each stellar
profile.  Nonzero-\(A\) cases are phenomenological BSk24-anchored effective
barotropes rather than BSk24 predictions; the experiment does not infer
microscopic composition, identify a phase transition, or establish a response
law independent of the baseline and feature geometry.

%% file: sections/02_framework.tex
\section{Method}
\label{sec:framework}

\subsection{Analytic representation of the unified BSk24 baseline}

The state variables are total energy density \(\epsilon\), including
rest-mass energy; isotropic pressure \(P\); and baryon number density
\(n_{\mathrm B}\).
The last quantity counts baryons per unit volume and is not interchangeable
with mass density \cite{ComposeManual2022,OBoyle2020}.  We use the unified
BSk24 EoS from the Brussels--Montreal family
\cite{Goriely2013HFB24,Pearson2018,Pearson2019Erratum,Perot2019,Perot2020}.
Its pressure is the published Appendix-C analytic representation of Pearson
et al., with the official implementation record used to check the
coefficients \cite{Pearson2018,PotekhinIoffe}.  The earlier analytic fit of Potekhin
et al.\ covers BSk19--BSk21 and is not used for the BSk24 coefficients
\cite{Potekhin2013Analytical}.

The published independent variable is mass density \(\rho\), mapped here to
total energy density by \(\epsilon=\rho c^2\) \cite{Pearson2018}.  No case is
extrapolated beyond
\begin{align}
10^6&\leq\rho/(\mathrm{g\,cm^{-3}})\leq2.69\times10^{15},\notag\\
5.6096\times10^{-7}&\leq
\epsilon/(\mathrm{MeV\,fm^{-3}})\leq1.509\times10^3\,.
\label{eq:retained-domain}
\end{align}
The upper value is the retained causal-domain endpoint of the undeformed
analytic baseline and is common to every case.

For density-based locations, we use the BSk24 saturation density
\(n_0=\BSkTwentyFourNSat\ \mathrm{fm^{-3}}\) \cite{Goriely2013HFB24}.  The
common map \(n_{B,0}(\epsilon)\) is obtained from the undeformed pressure in
Eq.~(C4) of Pearson et al.\ and normalized at the match state derived from
their Eq.~(C1).  It is not evaluated from Eq.~(C1) away from that anchor.  The
coordinate \(n_{\mathrm B,0}/n_0\) locates the same
input feature in every case; a pressure quoted at fixed \(n_{\mathrm B}\), by contrast,
uses the reconstructed density map of that case.

\subsection{Signed ramped-Gaussian deformation}

Pressure recovery by integrating a causal sound-speed representation is a
standard barotropic construction
\cite{Tews2018,Greif2019,Lindblom2018Causal}.  The profile used here is a
Gaussian multiplied by a smooth finite-width activation ramp:
\begin{align}
G(\epsilon)&=\exp\!\left[-\frac{(\epsilon-\epsilon_0)^2}{2\sigma^2}\right],
\label{eq:gaussian-profile}\\
W(x)&=\begin{cases}
0,&x\leq0,\\
10x^3-15x^4+6x^5,&0<x<1,\\
1,&x\geq1,
\end{cases}
\label{eq:smooth-window}\\
x&=\frac{\epsilon-\epsilon_{\rm match}}{\Delta},
\qquad \Phi(\epsilon)=G(\epsilon)W[x(\epsilon)],
\label{eq:applied-profile}\\
c_{\mathrm{s},A}^{2}(\epsilon)&=c_{\mathrm{s},0}^{2}(\epsilon)
+A\Phi(\epsilon).
\label{eq:deformation}
\end{align}
The quintic ramp joins its constant pieces with continuous first and second
derivatives.  The deformation is therefore exactly zero through
\(\epsilon_{\rm match}\), without a slope or curvature corner.  The only
varied input is the dimensionless signed coefficient \(A\): positive values
increase \(c_{\mathrm{s}}^2\), while negative values decrease it.

Because \(\max\Phi\neq1\), the realized peak change is
\begin{equation}
a_{\rm pk}=A\max_{\epsilon}\Phi(\epsilon),
\label{eq:peak-amplitude}
\end{equation}
not \(A\) itself.  The realized centroid and full width at half maximum
(FWHM) are defined by
\begin{align}
\epsilon_{\rm cent}
&=\frac{\int_{\epsilon_{\rm match}}^{\epsilon_{\rm end}}
\epsilon\Phi(\epsilon)\,d\epsilon}
{\int_{\epsilon_{\rm match}}^{\epsilon_{\rm end}}
\Phi(\epsilon)\,d\epsilon},\\
\Phi(\epsilon_-)&=\Phi(\epsilon_+)
=\tfrac12\max_{\epsilon}\Phi(\epsilon),
\qquad \mathrm{FWHM}=\epsilon_+-\epsilon_-.
\label{eq:realized-geometry}
\end{align}
Here \(\epsilon_{\rm end}\) is the common upper endpoint of the retained
analytic domain.  These definitions describe the applied profile rather than
its nominal Gaussian labels.  Its signed area is
\begin{equation}
\mathcal I_A=A\int_{\epsilon_{\rm match}}^{\epsilon_{\rm end}}
\Phi(\epsilon)\,d\epsilon,
\label{eq:realized-coordinates}
\end{equation}
with units of energy density.

\subsection{Thermodynamic reconstruction and checks}

Every case copies the undeformed effective state below the match.  At
\(\epsilon_{\rm match}=80\ \mathrm{MeV\,fm^{-3}}\), its shared anchor is
\begin{equation}
\begin{aligned}
n_{\rm match}&=\NMatch\ \mathrm{fm^{-3}},\\
P_{\rm match}&=\PMatch\ \mathrm{MeV\,fm^{-3}},\\
\mu_{\rm match}&=\MuMatch\ \mathrm{MeV}.
\end{aligned}
\label{eq:matched-state}
\end{equation}
These three values come from one BSk24 state and are not independent tuning
parameters.  Above the match,
\begin{align}
P_A(\epsilon)&=P_0(\epsilon_{\rm match})+
\int_{\epsilon_{\rm match}}^{\epsilon}c_{\mathrm{s},A}^2(\epsilon')\,d\epsilon',
\label{eq:pressure-reconstruction}\\
\Delta P_A(\epsilon)&=A\int_{\epsilon_{\rm match}}^{\epsilon}
\Phi(\epsilon')\,d\epsilon'.
\label{eq:pressure-memory}
\end{align}
The second expression is the integrated pressure offset.  Once the local
profile has decayed, it tends to \(\mathcal I_A\); we call this retained
offset \emph{pressure memory}.  The term is shorthand solely for this static
integral offset in the barotrope; it does not denote time dependence or
hysteresis.

For a zero-temperature effective one-fluid barotrope, the cold first law and
Euler relation are \cite{HaenselPotekhin2004,ComposeManual2022,OBoyle2020}
\begin{equation}
d\epsilon=\mu_{\mathrm{B}}^{\mathrm{eff}}\,d n_{\mathrm{B}},
\qquad \epsilon+P_A=\mu_{\mathrm{B}}^{\mathrm{eff}}n_{\mathrm{B}}.
\label{eq:first-law-euler}
\end{equation}
They give
\begin{align}
\ln\frac{n_{\mathrm{B}}(\epsilon)}{n_{\rm match}}
&=\int_{\epsilon_{\rm match}}^{\epsilon}
\frac{d\epsilon'}{\epsilon'+P_A(\epsilon')},
\label{eq:nb-reconstruction}\\
\mu_{\mathrm{B}}^{\mathrm{eff}}(\epsilon)
&=\frac{\epsilon+P_A}{n_{\mathrm{B}}},
\label{eq:mu-reconstruction}\\
\Gamma_{\mathrm{eq}}(\epsilon)
&=\frac{\epsilon+P_A}{P_A}c_{\mathrm{s},A}^2.
\label{eq:gamma-reconstruction}
\end{align}
Pressure is evaluated from the analytic BSk24 representation plus the closed
Gaussian--smootherstep primitive \cite{Pearson2018,Pearson2019Erratum}.  The
baryon state is reconstructed by cumulative Simpson integration in
\(\ln\epsilon\), anchored at the matched density.  Nonzero-\(A\) stellar
calculations use non-extrapolating log--log PCHIP representations for pressure
inversion and \(n_{\mathrm B}(\epsilon)\) \cite{FritschButland1984}, while
\(c_{\mathrm{s}}^2\) remains analytic at the recovered energy density.  The
\(A=0\) control uses the analytic BSk24 inverse directly.  Exact grids and
numerical settings are listed in Appendix~\ref{app:numerical-validation}.

Here \(\mu_{\mathrm{B}}^{\mathrm{eff}}\) is an effective baryon chemical
potential, not a species-resolved one, and
\(\Gamma_{\mathrm{eq}}=d\ln P/d\ln n_{\mathrm B}\) is the equilibrium
adiabatic index rather than a heat-capacity ratio \cite{ComposeManual2022}.

Before reconstruction, every value of the raw proposal on the complete
domain must be finite and satisfy
\begin{equation}
\epsilon>0,\qquad P_A>0,\qquad 0<c_{\mathrm{s},A}^2\leq1,
\label{eq:physical-gate}
\end{equation}
where the sound-speed bounds impose local mechanical stability and causality
\cite{Tews2018,BedaqueSteiner2015}.  A viable reconstruction must also join
the baseline continuously, preserve it exactly below the match, and have
positive, strictly increasing \(n_{\mathrm B}\) and positive
\(\mu_{\mathrm B}^{\mathrm{eff}}\).  A failed proposal is neither repaired nor
sent to a stellar calculation.  First-law and Euler residuals are recorded as
numerical diagnostics; no independent residual rejection threshold was
defined.  These admission and viability conditions are distinct from the
later fixed-mass support and retained-domain rules.  Neither the closure
residuals nor \(\Gamma_{\mathrm{eq}}\) establish microscopic or radial
stability \cite{KokkotasRuoff2001,Ibanez2013}.

Three extra quantities are recorded as warnings, not gates.  Besides the
pointwise comparison \(\Gamma_{\mathrm{eq}}<4/3\), we use the classical and
relativistic
fundamental derivatives
\begin{equation}
\mathcal G_{\rm C}=1+\frac12\frac{d\ln c_{\mathrm{s}}^2}{d\ln n_{\mathrm B}}
 +\frac12c_{\mathrm{s}}^2,
\qquad
\mathcal G_{\rm R}=\mathcal G_{\rm C}-\frac32c_{\mathrm{s}}^2.
\label{eq:fundamental-derivatives}
\end{equation}
They describe local convexity of the effective barotrope; a negative value
flags unusual wave-propagation behavior \cite{Ibanez2013}.  The value
\(4/3\) is a useful Newtonian reference, not a local general-relativistic
stability law \cite{Read2009}.  Relativistic radial stability is a global
eigenvalue problem \cite{KokkotasRuoff2001}.  None of these advisory checks
accepts, rejects, or repairs a case.

The derivative masks exclude domain endpoints and the match seam; resampling
is used only as a sensitivity check.  Their exact definitions, the distinct
closure masks, and all production settings appear in
Appendix~\ref{app:numerical-validation}.

\subsection{Stellar and tidal observables}

Accepted barotropes are mapped to cold, static, spherical, isotropic stars
with the Tolman--Oppenheimer--Volkoff equations
\cite{Tolman1939,OppenheimerVolkoff1939,ComposeManual2022}.  Here \(r\) is
circumferential radius, \(m(r)\) is enclosed gravitational mass, and the
central values are \(P_c=P(0)\) and \(\epsilon_c=\epsilon(0)\).  Because the
retained analytic fit starts at \(\rho=10^6\ \mathrm{g\,cm^{-3}}\), the
reported radius \(R\) and mass \(M=m(R)\) use that finite-pressure cutoff; no
zero-pressure extrapolation is added \cite{Pearson2018}.  Thus \(R\), \(M\),
\(k_2\), and \(\Lambda\) denote cutoff-defined observables, not demonstrated
vacuum-surface values.  The terminal tidal variable is supplied directly to
the standard exterior Love-number formula under the continuous-hadronic
convention.  The retained representation contains no declared physical
surface-density jump or discontinuity, so no jump correction is applied
\cite{TakatsyKovacs2020}.  This bookkeeping choice does not imply
\(\epsilon(R)=0\).  The one-sided cutoff study in
Appendix~\ref{app:numerical-validation} quantifies only in-domain sensitivity;
it neither supplies the missing lower-density layer nor determines the
correction to a physical \(P=0\) surface.  Absolute values retain this boundary
convention, while every matched comparison uses the same cutoff.

A fixed-mass result is reported only when the requested mass is bracketed on
the successful sampled sequence, ordered by central pressure, through its
sampled mass peak.  This is a numerical support rule, not independent
radial-mode-stability evidence.  A maximum mass is reported only when the
turning point of \(M(P_c)\) is bracketed and refined inside the retained EoS
domain; a highest sample or domain endpoint is not relabeled as \(M_{\max}\)
\cite{BardeenThorneMeltzer1966,Read2009,KastaunOhme2024}.

The tidal calculation uses compactness \(C=GM/(Rc^2)\), the quadrupolar Love
number \(k_2\), and the standard dimensionless tidal deformability
\cite{Hinderer2008,Hinderer2009Erratum,Postnikov2010,ComposeManual2022}
\begin{equation}
\Lambda=\frac{2}{3}k_2C^{-5}.
\label{eq:lambda}
\end{equation}
At fixed mass, this definition gives the exact bookkeeping identity
\begin{equation}
\Delta\ln\Lambda=\Delta\ln k_2+5\,\Delta\ln R.
\label{eq:lambda-decomposition}
\end{equation}
It does not separate independent physical mechanisms because \(k_2\) also
depends on compactness and on the complete interior profile.

\subsection{Fixed-mass response and radial support}

For any positive scalar observable \(X\), its fixed-mass response is
\begin{equation}
\mathcal R_X(A;M)=\ln\!\left[\frac{X_A(M)}{X_0(M)}\right].
\label{eq:response-function}
\end{equation}
Its exact fractional change is \(e^{\mathcal R_X}-1\).  For an exact saved
pair \(A=\pm0.8\), we compare equal deformation magnitudes directly rather
than infer a derivative at \(A=0\).  Here \(\mathcal R_X(0;M)=0\) exactly.

For \(A\neq0\), the radial diagnostic uses one threshold in density space for
every star.  At fraction \(q\) of the global realized EoS peak, its support is
\begin{equation}
\mathcal S_q(M;A)=\left\{r\in[0,R]:
|\Delta c_{\mathrm{s}}^2[\epsilon(r)]|\geq
q\max_{\epsilon}|\Delta c_{\mathrm{s}}^2(\epsilon)|
\right\}.
\label{eq:radial-support}
\end{equation}
The primary choice is \(q=1/2\); \(q=1/4\) and \(3/4\) test threshold
sensitivity.  A reached interval beginning at \(r=0\) is called
core-connected, two positive boundaries define a shell, and an empty set is
reported as not reached.  Boundaries are given in both \(r/R\) and enclosed
gravitational-mass fraction \(m(r)/M\); the latter is not a baryonic-mass
fraction.  At \(A=0\), the deformation and its threshold both vanish, so this
support classification is undefined and is reported as not applicable rather
than as the whole star.

%% file: sections/03_experiment.tex
\section{Numerical design}
\label{sec:experiment}

\subsection{One fixed geometry}

All calculations use
\begin{equation}
(\epsilon_{\rm match},\epsilon_0,\sigma,\Delta)
=(80,110,200,700)\ \mathrm{MeV\,fm^{-3}}.
\label{eq:geometry}
\end{equation}
Only \(A\) changes.  Exploratory calculations varied the match point, nominal
center, Gaussian width, and activation scale, including neighboring tests
around the adopted geometry.  We selected the displayed configuration as a
methodologically useful setting: it provides a broad two-sided admissible
amplitude range, retains fixed-mass solutions across the reporting range, and
places the realized feature within densities sampled by ordinary neutron
stars.  Selection followed inspection of exploratory thermodynamic and
stellar responses; it was neither preregistered nor a global optimization.
The geometry was then held fixed throughout the final amplitude campaign.
Its realized peak lies at
\(\RealizedPeakDensityOverNSat\,n_0\), within the range of
central densities sampled by ordinary neutron-star masses, while the match
remains below that region.  No preferred microscopic feature is implied.

\begin{table}[t]
\caption{Fixed design and final sampling.  The negative and positive counts
each include \(A=0\), which is shared in the combined total.}
\label{tab:controlled-experiment}
\centering
\small
\begin{tabular}{@{}p{0.39\columnwidth}p{0.53\columnwidth}@{}}
\toprule
Element & Value \\
\midrule
Baseline & Analytic representation of unified BSk24; exact \(A=0\) control \\
Geometry & \((80,110,200,700)\ \mathrm{MeV\,fm^{-3}}\) \\
Sampled amplitudes & \(\RawAmplitudeCount\) proposals: \(\RawAcceptedCount\) pass, \(\RawRejectedCount\) reject \\
Stellar amplitudes & \(\StellarAmplitudeCount\) unique; \(\NegativeStellarCount\) nonpositive and \(\PositiveStellarCount\) nonnegative \\
Fixed masses & \(0.2\)--\(2.0\,M_{\odot}\) in \(0.2\,M_{\odot}\) steps \\
Main endpoints & \(A=\SelectedNegativeAmplitude\) and \(+\SelectedPositiveAmplitude\); matched pair \(A=\pm\MatchedAmplitude\) \\
Realized peak/FWHM & \(\RealizedPeakDensityOverNSat\,n_0\); \(\RealizedFWHMLowerDensityOverNSat\)--\(\RealizedFWHMUpperDensityOverNSat\,n_0\) \\
\bottomrule
\end{tabular}
\end{table}

\begin{figure*}[!t]
  \centering
  \includegraphics[width=0.98\linewidth]{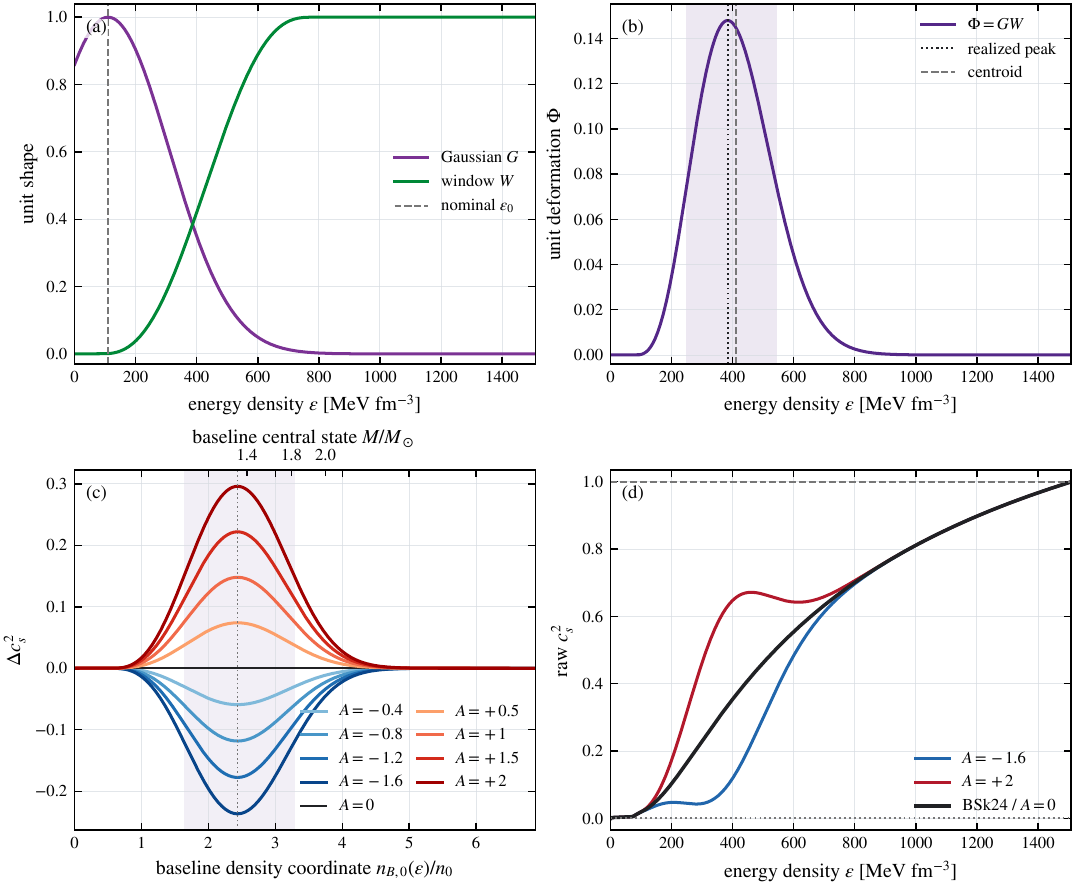}
  \caption{Realized deformation geometry.  Panel (a) shows the Gaussian
  \(G\) and activation ramp \(W\).  Panel (b) shows their product \(\Phi\),
  with its peak, centroid, and FWHM\@.  Panel (c) places representative signed
  changes on the undeformed BSk24 density coordinate; the upper marks show
  baseline central states at \(1.4\), \(1.8\), and \(2.0\,M_{\odot}\).  Panel
  (d) shows the raw squared sound speed at the two selected reporting
  endpoints and at \(A=0\).}
  \label{fig:realized-deformation}
\end{figure*}

Figure~\ref{fig:realized-deformation} shows what the nominal parameters
actually produce.  Because the activation ramp is still small at
\(\epsilon_0\), the product \(\Phi=GW\) peaks at
\(\RealizedPeakEnergy\ \mathrm{MeV\,fm^{-3}}\), not at
\(110\ \mathrm{MeV\,fm^{-3}}\).  Its centroid is
\(\RealizedCentroid\ \mathrm{MeV\,fm^{-3}}\), and its FWHM is
\(\RealizedFWHM\ \mathrm{MeV\,fm^{-3}}\).

\subsection{Admission and comparison protocol}

Because \(\Phi\geq0\), the complete-domain sound-speed gate gives
\begin{equation}
A_{\min}=\max_{\Phi>0}\!\left(-\frac{c_{\mathrm{s},0}^2}{\Phi}\right),
\qquad
A_{\max}=\min_{\Phi>0}\!\left(\frac{1-c_{\mathrm{s},0}^2}{\Phi}\right).
\label{eq:amplitude-bounds}
\end{equation}
The admissible set is \(A_{\min}<A\leq A_{\max}\): the lower endpoint would
reach \(c_{\mathrm{s}}^2=0\), whereas the upper endpoint reaches the allowed
value \(c_{\mathrm{s}}^2=1\).  Analytic interval membership is kept distinct
from the floating-point evaluation of each proposal.  Advisory values of
\(\Gamma_{\mathrm{eq}}\), \(\mathcal G_{\rm C}\), and
\(\mathcal G_{\rm R}\) have no admission authority.

The final comparison uses fixed gravitational masses from \(0.2\) to
\(2.0\,M_\odot\) in \(0.2\,M_\odot\) increments.  The inward reporting
endpoints are \(A=-1.6\) and \(+2\), and \(A=\pm0.8\) supplies the
equal-magnitude sign comparison.  These amplitudes are saved campaign points;
no interpolation between amplitudes or target masses enters the displayed
results.

%% file: sections/04_results.tex
\section{Results}
\label{sec:results}

Unless stated otherwise, \(A=-1.6\) and \(+2\) are the selected inward
reporting endpoints.  The exact pair \(A=\pm0.8\) is the equal-magnitude
comparison.

\subsection{Amplitude admission and control}

The numerically resolved complete-domain interval is
\begin{equation}
\begin{aligned}
A_{\min}&<A\leq A_{\max},\\
A_{\min}&\simeq\FormalAmplitudeLower,\qquad
A_{\max}\simeq\FormalAmplitudeUpper.
\end{aligned}
\label{eq:hard-amplitude-interval}
\end{equation}
The lower limiter reaches \(c_{\mathrm{s}}^2=0\) near
\(\FormalAmplitudeLowerEnergy\ \mathrm{MeV\,fm^{-3}}\), whereas the upper
limiter reaches \(c_{\mathrm{s}}^2=1\) near
\(\FormalAmplitudeUpperEnergy\ \mathrm{MeV\,fm^{-3}}\).  The rounded values
in Eq.~\eqref{eq:hard-amplitude-interval} are for display; the saved campaign
uses the full-precision bounds.

The sampled map contains \(\RawAcceptedCount\) proposals that satisfy the
floating-point physical checks and \(\RawRejectedCount\) rejected outward
samples.  One passing row is the exact, mathematically open lower endpoint,
where roundoff leaves a minimum \(c_{\mathrm{s}}^2\) of
\(8.3\times10^{-15}\); it is retained only as a boundary diagnostic.  The
remaining \(\StellarAmplitudeCount\) formal-interval members are eligible for
stellar calculations.  The exact upper endpoint is eligible.  The four
outward samples fail through nonpositive \(c_{\mathrm{s}}^2\) on the softened
side or superluminal \(c_{\mathrm{s}}^2\) on the stiffened side.  No accepted
proposal is clipped, repaired, or extrapolated.

The advisory onsets remain separate from admission.  On the softened side,
the last unflagged and first flagged samples are \(-1.675,-1.6875\) for
\(\Gamma_{\mathrm{eq}}<4/3\), and \(-1.75,-1.7625\) for both
\(\mathcal G_{\rm C}<0\) and \(\mathcal G_{\rm R}<0\).  On the stiffened
side, \(\mathcal G_{\rm R}<0\) first appears between \(2.45\) and
\(2.4625\); the other warnings are not reached before \(A_{\max}\).  These
are sampled brackets, not exact transitions or rejection limits.  The selected
endpoints lie inward of every sampled warning onset.

\begin{figure*}[!t]
  \centering
  \includegraphics[width=0.96\linewidth]{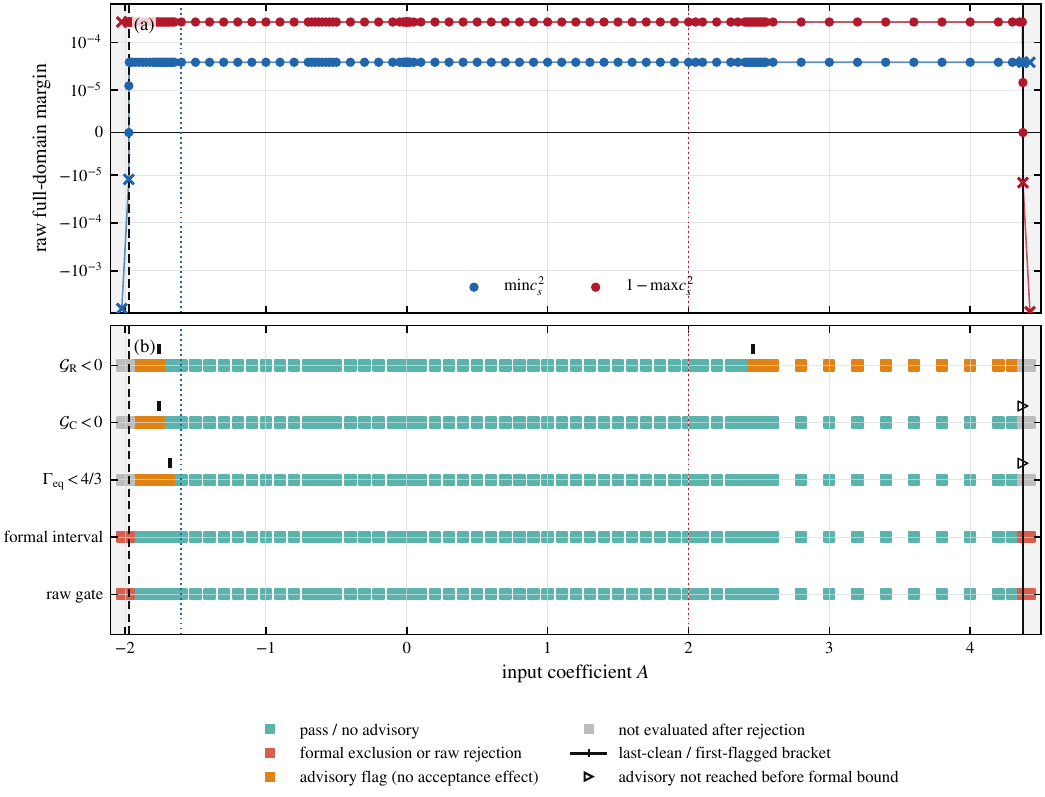}
  \caption{Amplitude admission and advisory map.  Panel (a) shows the raw
  full-domain margins \(\min c_{\mathrm{s}}^2\) and
  \(1-\max c_{\mathrm{s}}^2\); crosses are rejected outward samples.  Panel
  (b) separates the floating-point physical checks, membership in the formal
  open/closed interval, and the three advisory flags.  Short black bars join
  the last unflagged and first flagged saved samples; open arrows mean that no
  advisory onset was reached before the formal bound.  Black lines mark the
  formal interval, and blue and red dotted lines mark \(A=-1.6,+2\).}
  \label{fig:amplitude-admission}
\end{figure*}

Figure~\ref{fig:amplitude-admission} separates the physical checks, formal
interval, advisory quantities, and reporting choices.

The direct analytic baseline and the reconstructed \(A=0\) control are
array-identical for the retained thermodynamic fields and saved \(M\), \(R\),
and \(\Lambda\) sequences.  For orientation, the cutoff-defined control at
\(1.4\,M_\odot\) gives \(R_{1.4}=12.602\ \mathrm{km}\),
\(k_{2,1.4}=0.09343\), and \(\Lambda_{1.4}=524.4\).
Appendix~\ref{app:numerical-validation} compares this point with published
BSk24 calculations and explains why the comparison is a surface-convention
consistency check rather than a precision tidal benchmark.

\subsection{Thermodynamic response}

\begin{figure*}[!t]
  \centering
  \includegraphics[width=0.98\linewidth]{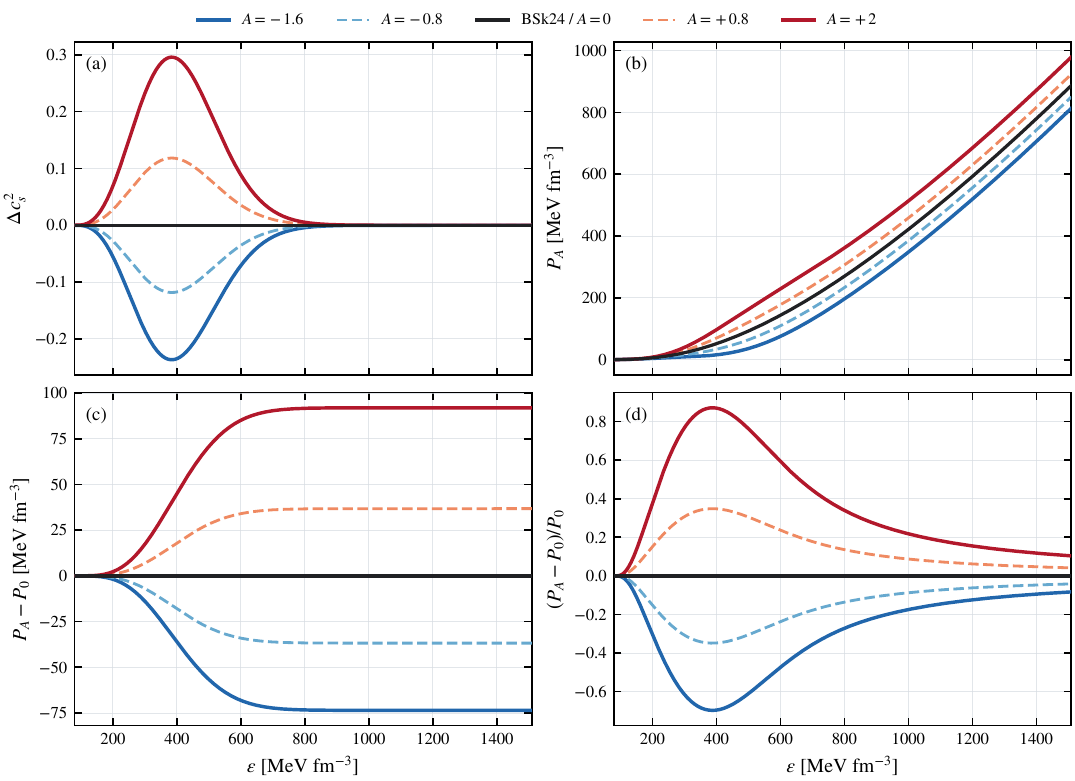}
  \caption{BSk24-anchored effective barotropes over the part of the retained domain with
  \(\epsilon\geq80\ \mathrm{MeV\,fm^{-3}}\).  Panel (a) shows the local input
  \(\Delta c_{\mathrm{s}}^2\); panel (b) shows the resulting absolute
  \(P_A(\epsilon)\) curves; panels (c,d) show the absolute and fractional
  pressure shifts relative to \(A=0\).  Dashed curves are the moderate pair
  \(A=\pm0.8\); solid colored curves are the reporting endpoints
  \(A=\SelectedNegativeAmplitude,+\SelectedPositiveAmplitude\).  Every curve
  is a saved, accepted profile evaluated at its displayed amplitude; no
  interpolation between amplitudes is used.}
  \label{fig:thermodynamic-propagation}
\end{figure*}

Figure~\ref{fig:thermodynamic-propagation} shows the undeformed barotrope,
the equal-magnitude pair \(A=\pm0.8\), and the two reporting endpoints.
The absolute \(P_A(\epsilon)\) curves make the deformed EoSs visible directly:
softening lowers pressure at fixed energy density, while stiffening raises it.
The realized endpoint peak changes are
\(\NegativePeakDeltaCsTwo\) and \(+\PositivePeakDeltaCsTwo\).  After the
local sound-speed feature has decayed, the absolute pressure shifts saturate
near
\(\NegativePressureMemory\ \mathrm{MeV\,fm^{-3}}\) and
\(+\PositivePressureMemory\ \mathrm{MeV\,fm^{-3}}\), while their fractional
importance decreases because the baseline pressure continues to rise.  This
is the pressure memory defined in Eq.~\eqref{eq:pressure-memory}.

\subsection{Fixed-mass response}

\begin{figure*}[!t]
  \centering
  \includegraphics[width=0.98\linewidth]{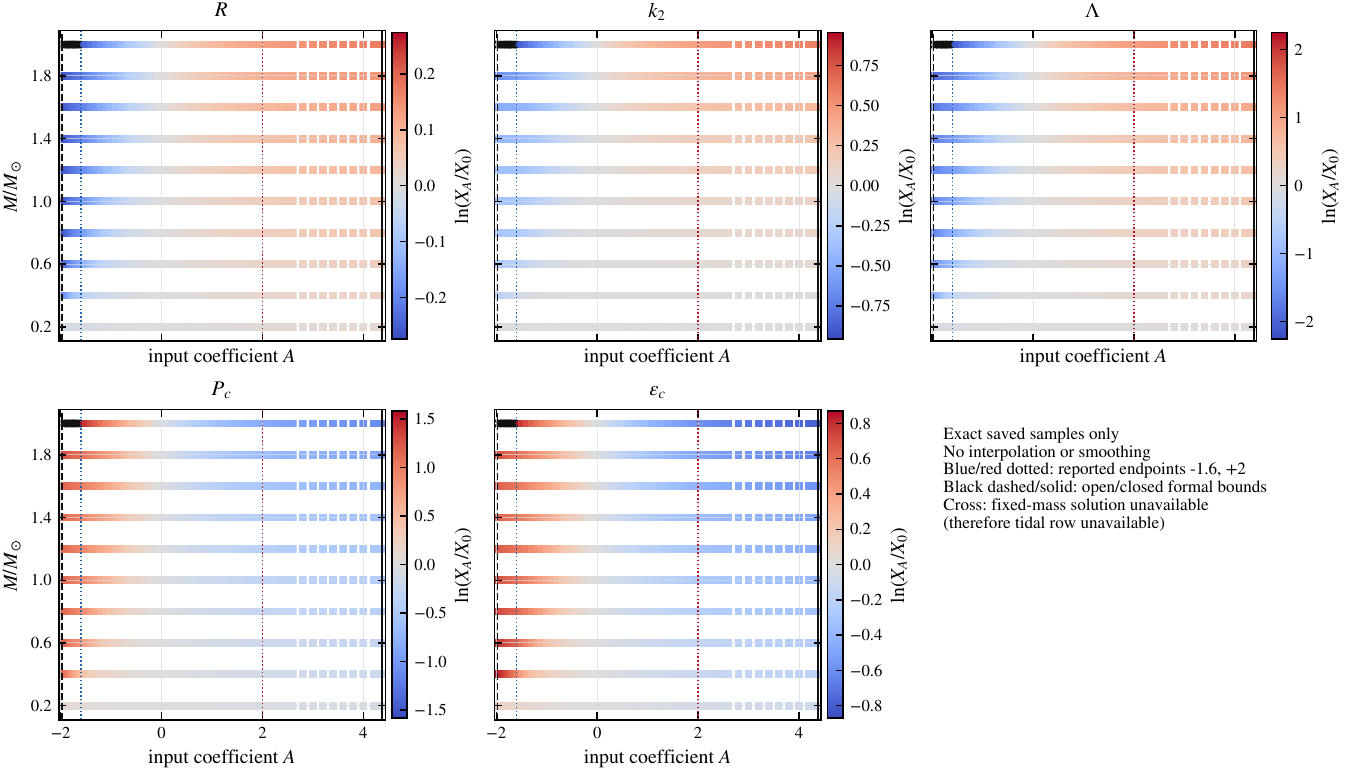}
  \caption{Saved fixed-mass response map.  Color gives
  \(\mathcal R_X=\ln(X_A/X_0)\) for \(R\), \(k_2\), \(\Lambda\), \(P_c\),
  and \(\epsilon_c\).  Black dashed and solid lines are the open lower and
  closed upper formal bounds; blue and red dotted lines mark the reporting
  endpoints.  Crosses are unavailable fixed-mass rows, which also make the
  corresponding tidal row unavailable.}
  \label{fig:fixed-mass-response}
\end{figure*}

Figure~\ref{fig:fixed-mass-response} replaces overlapping response curves
with a two-dimensional map.  Each square is one saved fixed-mass result from
the tightest ordinary-differential-equation (ODE) tolerance stage.  No
interpolation or smoothing between displayed amplitudes or target masses is
used to fill the map.  Stiffening raises \(P_A(\epsilon)\), so a fixed
gravitational mass is supported at lower \(P_c\) and \(\epsilon_c\), with
larger \(R\) and lower compactness \(C\); softening gives the opposite
response.  Compactness and complete-profile changes both propagate into
\(k_2\), and \(\Lambda=(2/3)k_2C^{-5}\) amplifies the compactness dependence.
These effects are coupled, so no fractional response is assigned uniquely to
one mechanism.  The color scale is separate for each observable so that the
sign and mass dependence remain readable.

At \(1.4\,M_{\odot}\), the endpoint change in \(\Lambda\) is
\(\NegativeLambdaPercentOneFour\%\) for softening and
\(+\PositiveLambdaPercentOneFour\%\) for stiffening.  At
\(2.0\,M_{\odot}\), the
corresponding stress-endpoint changes are
\(\NegativeLambdaPercentTwoZero\%\) and
\(+\PositiveLambdaPercentTwoZero\%\).  These large endpoint responses are not used
as a small-amplitude approximation.  Crosses at \(2.0\,M_{\odot}\) and
\(A\leq-1.65\) mean that the fixed mass was not bracketed on the saved
successful sampled sequence through its mass peak; they do not represent zero
response.

Table~\ref{tab:matched-amplitude-response} gives the exact pair
\(A=\pm0.8\).  The softened member needs a larger rise in \(P_c\) and
\(\epsilon_c\) than the stiffened member needs in the opposite direction.
The response is mass dependent and not mirror symmetric.
\mbox{The saved pair displays both properties.}

\input{tables/matched_sign_A_pm_0p8.tex}

\subsection{Stellar sequences and domain reach}

\begin{figure*}[!t]
  \centering
  \includegraphics[width=0.94\linewidth]{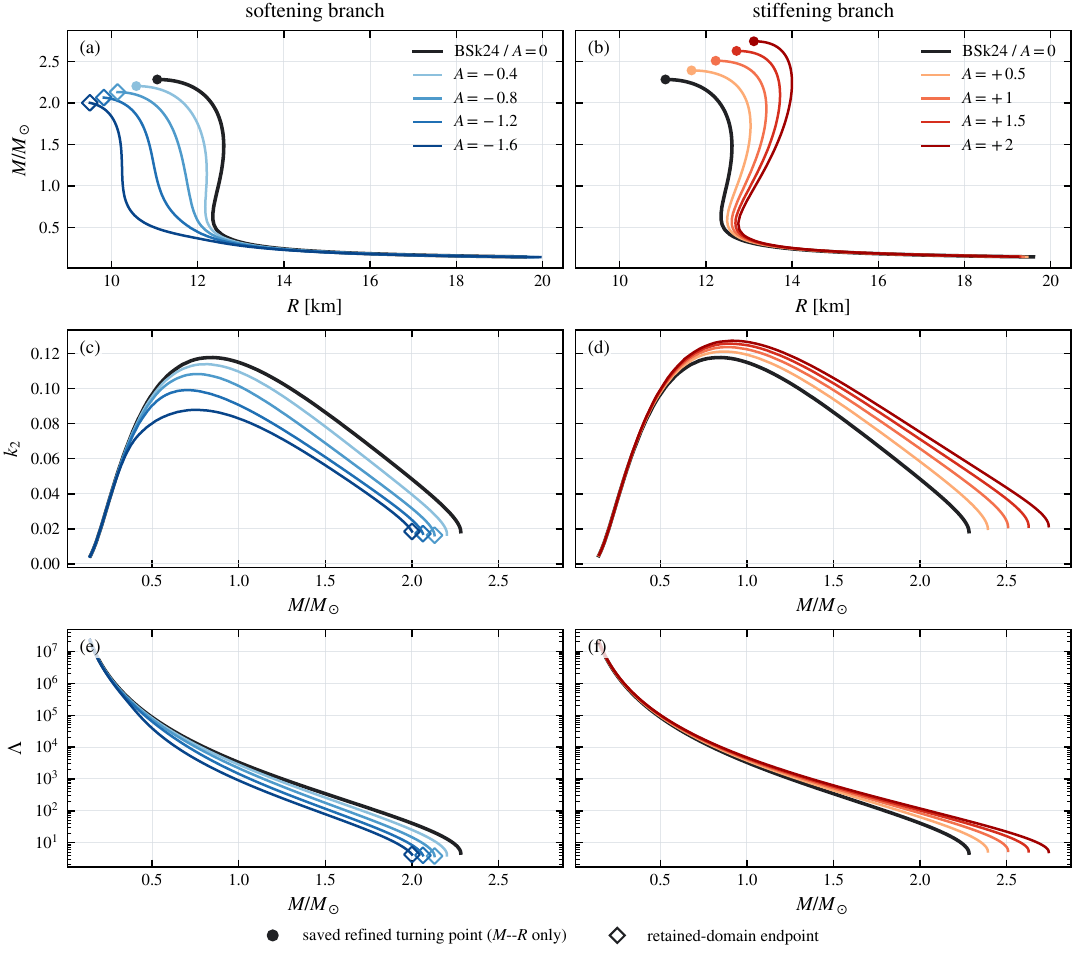}
  \caption{Representative successful pre-peak stellar sequences from exact saved points.
  Columns separate softening and stiffening; rows show mass--radius,
  \(k_2(M)\), and \(\Lambda(M)\).  Filled circles mark a refined turning point
  in the mass--radius panels only.  Open diamonds mark a retained-domain
  endpoint, explicitly not \(M_{\max}\).}
  \label{fig:stellar-sequences}
\end{figure*}

Figure~\ref{fig:stellar-sequences} shows representative successful pre-peak
sequence segments.  Filled circles in the mass--radius row are refined turning points.
Open diamonds are saved endpoints of the retained EoS domain and are not
maximum masses.  No line bridges a failed gap or continues outside the
analytic domain.

For this selected fixed geometry, softening lowers the pressure support and
pushes a fixed-mass star toward higher central density.  The successful
sequence can therefore exhaust the retained EoS domain before a turning point
is established.  Stiffening instead adds pressure support, lowers the required
central state, and extends the successful branch.  These domain and
turning-point outcomes remain distinct.

At \(A=-1.6\), the bracketed \(2.0\,M_{\odot}\) star has
\(\epsilon_c=\EpsCNegativeTwoZero\ \mathrm{MeV\,fm^{-3}}\), about
\(4\ \mathrm{MeV\,fm^{-3}}\) below the
retained endpoint.  It is a valid saved fixed-mass solution but a near-domain
stress test.  The sequence itself reaches
\(M_{\rm end}=2.0004\,M_{\odot}\) without a turning point.  We therefore
report no
\(M_{\max}\) for that case.

\subsection{Where the feature sits inside a star}

\begin{figure*}[!t]
  \centering
  \includegraphics[width=0.98\linewidth]{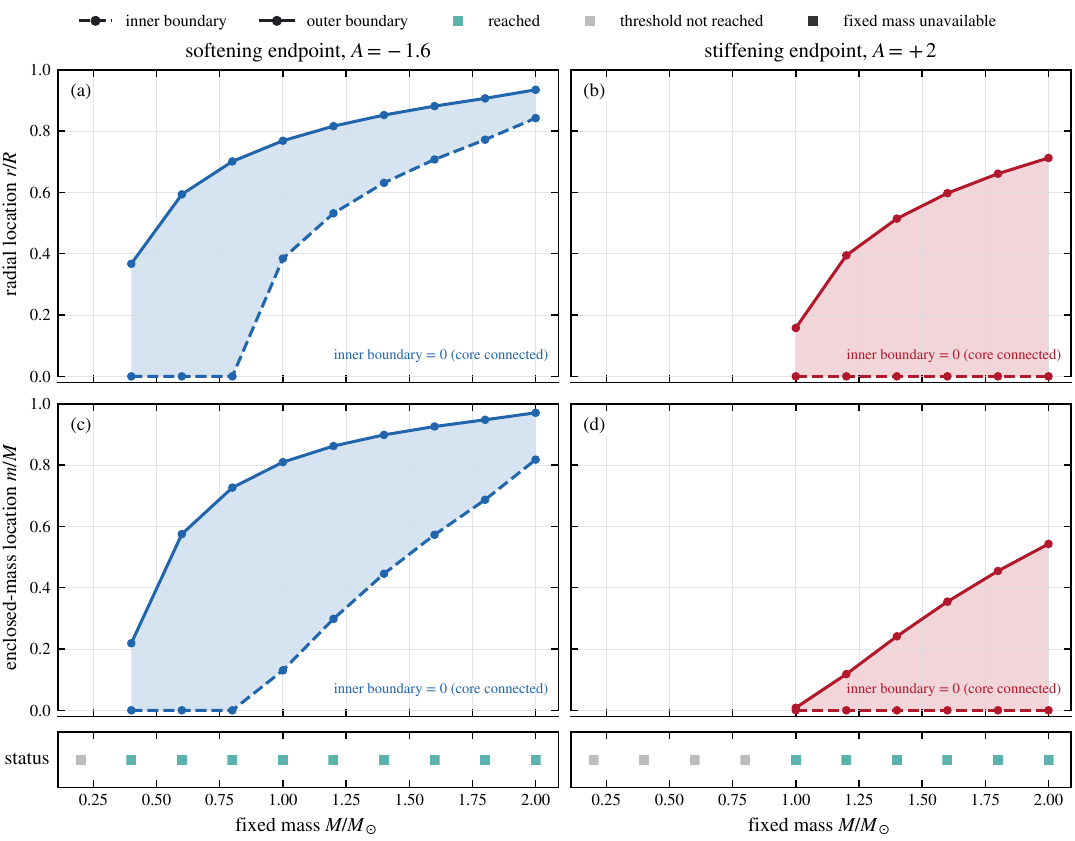}
  \caption{Endpoint support above 50 per cent of the global realized
  deformation peak.  Columns show \(A=-1.6\) and \(+2\); the upper rows give
  \(r/R\) and enclosed gravitational-mass fraction \(m(r)/M\).  Dashed and
  solid curves are actual inner and outer boundaries only.  A zero inner
  boundary means core connected.  The thin status strip separately records
  reached, not reached, and unavailable fixed-mass cases, so a missing
  physical boundary is never drawn at zero.}
  \label{fig:radial-support}
\end{figure*}

Figure~\ref{fig:radial-support} shows the support above 50 per cent of the
global realized EoS peak.  This normalized support is the fixed half-maximum
energy-density interval
\(\epsilon_-\simeq248.6\) to
\(\epsilon_+\simeq544.8\ \mathrm{MeV\,fm^{-3}}\).  Hydrostatic balance gives
\(dP/dr<0\), while the admitted barotropes have
\(dP/d\epsilon=c_{\mathrm{s}}^2>0\); hence
\(d\epsilon/dr=(dP/dr)/c_{\mathrm{s}}^2<0\).  The central density therefore
sets the topology directly:
\begin{equation}
\begin{cases}
\epsilon_c<\epsilon_- &: \text{threshold not reached},\\
\epsilon_-\leq\epsilon_c\leq\epsilon_+ &: \text{core connected},\\
\epsilon_c>\epsilon_+ &: \text{off-center shell}.
\end{cases}
\label{eq:radial-support-classification}
\end{equation}

At \(A=+2\), \(\epsilon_c\) rises from \(232.6\) to
\(251.8\ \mathrm{MeV\,fm^{-3}}\) between \(0.8\) and
\(1.0\,M_{\odot}\), crossing \(\epsilon_-\); the threshold is first reached
at \(1.0\,M_{\odot}\) and remains core connected through
\(2.0\,M_{\odot}\), where \(\epsilon_c=362.4\ \mathrm{MeV\,fm^{-3}}\).
At that mass its outer boundary is
\(r/R=\PositiveSupportOuterRadiusTwoZero\) and encloses gravitational-mass
fraction \(m/M=\PositiveSupportOuterMassTwoZero\).

At \(A=-1.6\), \(\epsilon_c=180.9\ \mathrm{MeV\,fm^{-3}}\) at
\(0.2\,M_{\odot}\), below \(\epsilon_-\), and
\(366.8\ \mathrm{MeV\,fm^{-3}}\) at \(0.4\,M_{\odot}\), inside the interval.
The change from \(540.4\) to \(595.7\ \mathrm{MeV\,fm^{-3}}\) between
\(0.8\) and \(1.0\,M_{\odot}\) crosses \(\epsilon_+\), explaining the
transition from a core-connected region to an off-center shell.  At
\(2.0\,M_{\odot}\), the shell spans
\(r/R=\NegativeSupportInnerRadiusTwoZero\)--\(\NegativeSupportOuterRadiusTwoZero\)
and \(m/M=\NegativeSupportInnerMassTwoZero\)--\(\NegativeSupportOuterMassTwoZero\).

The feature is fixed in density space; the star readjusts relative to it.
Stiffening raises pressure support and lowers the central state required at
fixed mass, whereas softening raises the required central density and can move
the center beyond the fixed interval.  This construction diagnoses the strong
local \(\Delta c_{\mathrm{s}}^2\) region, not the full pressure response or a
composition or phase boundary: integrated pressure memory persists after the
local feature decays.  The all-amplitude and 25/50/75 per cent extension
appears in Fig.~\ref{fig:radial-support-thresholds}.

%% file: tables/matched_sign_A_pm_0p8.tex
\begin{table*}[t]
\caption{Equal-magnitude comparison at the exact saved pair $A=\pm0.8$.
Entries are percentage changes from the shared exact $A=0$ baseline.  The
last column gives the central energy density as a fraction of the retained
EoS endpoint.}
\label{tab:matched-amplitude-response}
\centering
\begin{tabular}{ccrrrrrr}
\toprule
$M/M_\odot$ & $A$ & $\Delta R/R_0$ [\%] & $\Delta k_2/k_{2,0}$ [\%] & $\Delta\Lambda/\Lambda_0$ [\%] & $\Delta P_c/P_{c,0}$ [\%] & $\Delta\epsilon_c/\epsilon_{c,0}$ [\%] & $\epsilon_c/\epsilon_{\rm end}$ \\
\midrule
1.4 & $-0.8$ & -7.89 & -18.60 & -46.02 & +59.10 & +35.73 & 0.3680 \\
1.4 & $+0.8$ & +4.11 & +10.67 & +35.33 & -21.73 & -17.33 & 0.2241 \\
\addlinespace
1.8 & $-0.8$ & -9.67 & -26.74 & -55.93 & +77.01 & +38.70 & 0.4916 \\
1.8 & $+0.8$ & +5.98 & +20.35 & +60.92 & -30.84 & -22.94 & 0.2731 \\
\addlinespace
2.0 & $-0.8$ & -11.05 & -34.51 & -63.52 & +101.0 & +46.42 & 0.6208 \\
2.0 & $+0.8$ & +7.50 & +30.40 & +87.17 & -37.95 & -27.07 & 0.3092 \\
\bottomrule
\end{tabular}
\end{table*}

%% file: sections/05_scope_conclusions.tex
\section{Discussion}
\label{sec:scope}

The causal chain is
\(\Delta c_{\mathrm{s}}^2\to\Delta P_A(\epsilon)\to\) central-state
relocation \(\to\) changes in \(R\), compactness, \(k_2\), and \(\Lambda\).
Integration makes the pressure step nonlocal; at fixed mass, the relocated
central state and pressure profile alter both compactness and the complete
interior profile entering \(k_2\).  The effects are coupled, so no fractional
response is assigned uniquely to one mechanism.

At finite amplitude, the nonlinear reconstruction factor
\((\epsilon+P)^{-1}\) helps make the equal-magnitude response non-mirror
symmetric under \(A\mapsto-A\).  The radial classification shows where the
same fixed density-space feature is sampled inside a readjusted star, not a
composition boundary or phase transition.  Its high-mass core--shell
distinction persists under the tested thresholds, whereas the inferred onset
mass remains dependent on threshold, geometry, and mass sampling.

All reported stellar observables use the finite in-domain surface convention
defined in Sec.~\ref{sec:framework}.  The one-sided cutoff study from
\(10^6\) to \(10^8\,\mathrm{g\,cm^{-3}}\), summarized in
Table~\ref{tab:numerical-validation} and
Fig.~\ref{fig:numerical-validation}, quantifies sensitivity within the
supported analytic domain.  Raising the cutoff to
\(10^8\,\mathrm{g\,cm^{-3}}\) changed \(R\) by at most \(0.0309\) km,
\(k_2\) by \(1.154\times10^{-3}\) (\(1.235\%\)), and \(\Lambda\) by
\(0.648\) ppm relative to the \(10^6\,\mathrm{g\,cm^{-3}}\) result.  This
test does not supply the missing lower-density layer or determine the
correction to a physical \(P=0\) surface
\cite{Servignat2024,Davis2024UnifiedCrust}.  Because every amplitude uses the
same cutoff, the matched relative responses are the principal comparison;
absolute \(R\), \(k_2\), and \(\Lambda\) values remain tied to the stated
boundary convention.

Because the fixed geometry was selected after outcome-informed exploration
rather than a formal global optimization, the numerical response should not
be generalized across feature geometries.

The numerical values belong to the analytic BSk24 baseline, its
phenomenological BSk24-anchored effective barotropes, the selected
ramped-Gaussian geometry, and the retained domain.  The reconstruction supplies
no particle fractions, species chemical potentials, finite-temperature
properties, or transport coefficients, and the imposed feature is not
evidence for a microscopic constituent or phase transition
\cite{Mroczek2023}.  The stellar calculation assumes general relativity,
spherical symmetry, isotropic pressure, and negligible rotation
\cite{Tolman1939,OppenheimerVolkoff1939,Hinderer2008}.  No independently
implemented tidal-solver benchmark or radial-mode calculation was performed;
the tidal results are supported by the internal refinement, bookkeeping, and
convention-limited comparison reported in
Appendix~\ref{app:numerical-validation}.  Equilibrium and frozen-composition
perturbations need not place the radial-stability boundary at the same point
\cite{CanullanPascual2025}; the effective one-fluid reconstruction used here
cannot select a reaction-timescale regime.

\section{Conclusions}
\label{sec:conclusions}

Holding the analytic BSk24 baseline, feature geometry, and comparison protocol
fixed, we reconstructed signed localized sound-speed changes and compared
their stellar consequences at fixed gravitational mass.  A local change in
\(dP/d\epsilon\) changes the integrated pressure, relocates the central state,
and thereby produces mass-dependent and sign-asymmetric changes in \(R\),
\(k_2\), and \(\Lambda\).  For the selected endpoints \(A=-1.6\) and \(+2\),
the radial audit distinguishes a high-mass softened shell from a
core-connected stiffened feature.  Same-cutoff relative responses are the
principal result because the absolute observables retain the finite-domain
surface convention.  These findings apply to the selected BSk24-anchored
effective barotropes and are not BSk24 predictions, universal response laws,
or microscopic phase claims.

\begin{acknowledgments}
The author thanks Charalampos Moustakidis and Theodoros Diakonidis for
insightful suggestions concerning the numerical experiments and thermodynamic
diagnostics used to assess the effective barotropes and identify
potential physical violations.  The author received no external funding for
this research.
\end{acknowledgments}

\paragraph*{Author contributions.}
The author conceived the study, developed and implemented the method,
performed the calculations, analyzed the results, prepared the figures and
data products, and wrote and revised the manuscript.

\paragraph*{Competing interests.}
The author declares no competing interests.

\paragraph*{Use of artificial intelligence tools.}
During manuscript preparation, the author used OpenAI Codex (GPT-5, accessed
in August 2026) for language editing, LaTeX preparation, bibliography review,
and consistency checks.  The
tool was not used to generate or modify the scientific calculations or their
results.  The author reviewed all suggested changes and takes full
responsibility for the manuscript and its scientific content.

\paragraph*{Data and code availability.}
The maintained implementation is publicly available under the MIT License at
\url{https://github.com/PapathanasiouIoannis/EoS-generation}.  Analysis-ready
derived data underlying the displayed figures and tables, the exact campaign
declaration and registry, a sanitized checksum-pinned source capsule, and
verification notes are archived in the immutable campaign-data release
\cite{Papathanasiou2026CampaignData}.
The original calculation packets remain privately retained because their
operational metadata contain machine-specific paths; they are available from
the author on reasonable request.  The LaTeX source accompanying this preprint
is distributed by arXiv.  The release documentation records the sanitization
transformations and the limits of byte-identical rerunning.

%% file: sections/appendices.tex
\appendix

\section{Closed pressure primitive}
\label{app:pressure-primitive}

The pressure change is the definite integral of the imposed sound-speed
profile, so it can be evaluated without numerical quadrature.  On the ramp
interval, let
\begin{align}
x&=\frac{\epsilon-\epsilon_{\rm match}}{\Delta},
&z&=\epsilon-\epsilon_0,\nonumber\\
W(x)&=10x^3-15x^4+6x^5.
\end{align}
Writing \(W=\sum_{k=0}^{5}b_k z^k\) after the shift to \(z\), define
\begin{align}
F_{\rm ramp}(\epsilon)
&=\sum_{k=0}^{5}b_k\,[I_k(z)-I_k(z_{\rm match})],\notag\\
\epsilon_{\rm r}&=\epsilon_{\rm match}+\Delta,
&z_{\rm r}&=\epsilon_{\rm r}-\epsilon_0\,.
\end{align}
For compactness, let \(F_{\rm r}=F_{\rm ramp}(\epsilon_{\rm r})\).  The
complete pressure contribution is then
\begin{equation}
P_A-P_0=A
\begin{cases}
0, & \epsilon\leq\epsilon_{\rm match},\\
F_{\rm ramp}(\epsilon),
 & \epsilon_{\rm match}<\epsilon<\epsilon_{\rm r},\\
F_{\rm r}+I_0(z)-I_0(z_{\rm r}),
 & \epsilon\geq\epsilon_{\rm r}.
\end{cases}
\label{eq:exact-pressure-primitive}
\end{equation}
Here the last line is the completed ramp area plus the fully activated Gaussian
tail.  The Gaussian moments follow from
\begin{align}
I_0(z)&=\sigma\sqrt{\frac{\pi}{2}}\,
 \operatorname{erf}\!\left(\frac{z}{\sqrt{2}\sigma}\right),\nonumber\\
I_1(z)&=-\sigma^2e^{-z^2/(2\sigma^2)},\nonumber\\
I_k(z)&=-\sigma^2z^{k-1}e^{-z^2/(2\sigma^2)}\notag\\
&\quad +(k-1)\sigma^2I_{k-2}(z),\qquad k\geq2.
\label{eq:gaussian-moment-recurrence}
\end{align}
The coefficients \(b_k\) are fixed uniquely by the displayed polynomial and
the shift; no fitted coefficients enter this calculation.  Definite
differences enforce the pressure match at \(\epsilon_{\rm match}\).  The
independent quadrature check is reported in
Table~\ref{tab:numerical-validation}.  The published analytic BSk24
representations and their coefficient mapping are used as documented in the
source code and original references rather than reproduced here
\cite{Pearson2018,Pearson2019Erratum,PotekhinIoffe}.

\section{Auxiliary response diagnostics}
\label{app:auxiliary-response}

\begin{figure*}[!t]
  \centering
  \includegraphics[width=0.90\textwidth]{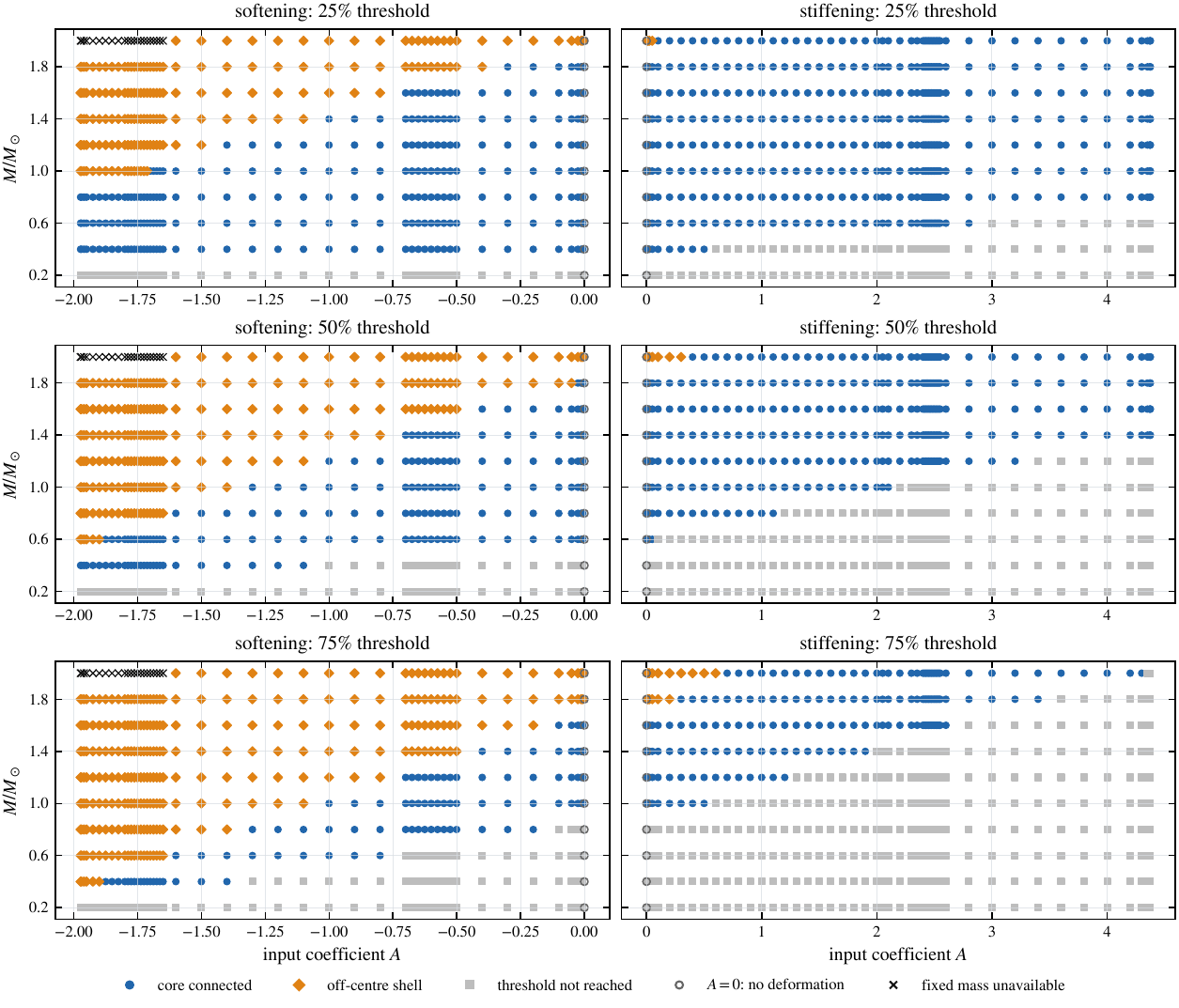}
  \caption{Radial classification over the complete saved amplitude--mass
  map.  Columns separate signs and rows use 25, 50, and 75 per cent of the
  global realized deformation peak.  Blue circles are core connected, orange
  diamonds are off-center shells, and gray squares mean the threshold is not
  reached.  Open circles at \(A=0\) mark a zero-deformation diagnostic that is
  not applicable; crosses mark an unavailable fixed mass.  Every marker is a
  saved case; no interpolation between displayed amplitudes or target masses
  is used.}
  \label{fig:radial-support-thresholds}
\end{figure*}

The radial support classification is not tied to one contour or only to the
two reporting endpoints.  Figure~\ref{fig:radial-support-thresholds} shows all
stellar-eligible amplitudes and all ten fixed masses at 25, 50, and 75 per
cent of the global realized EoS peak.  The high-mass softened shell and
stiffened core connection persist, although the mass at which a threshold is
first reached changes with both \(A\) and the threshold.

\begin{figure*}[!t]
  \centering
  \includegraphics[width=0.90\textwidth]{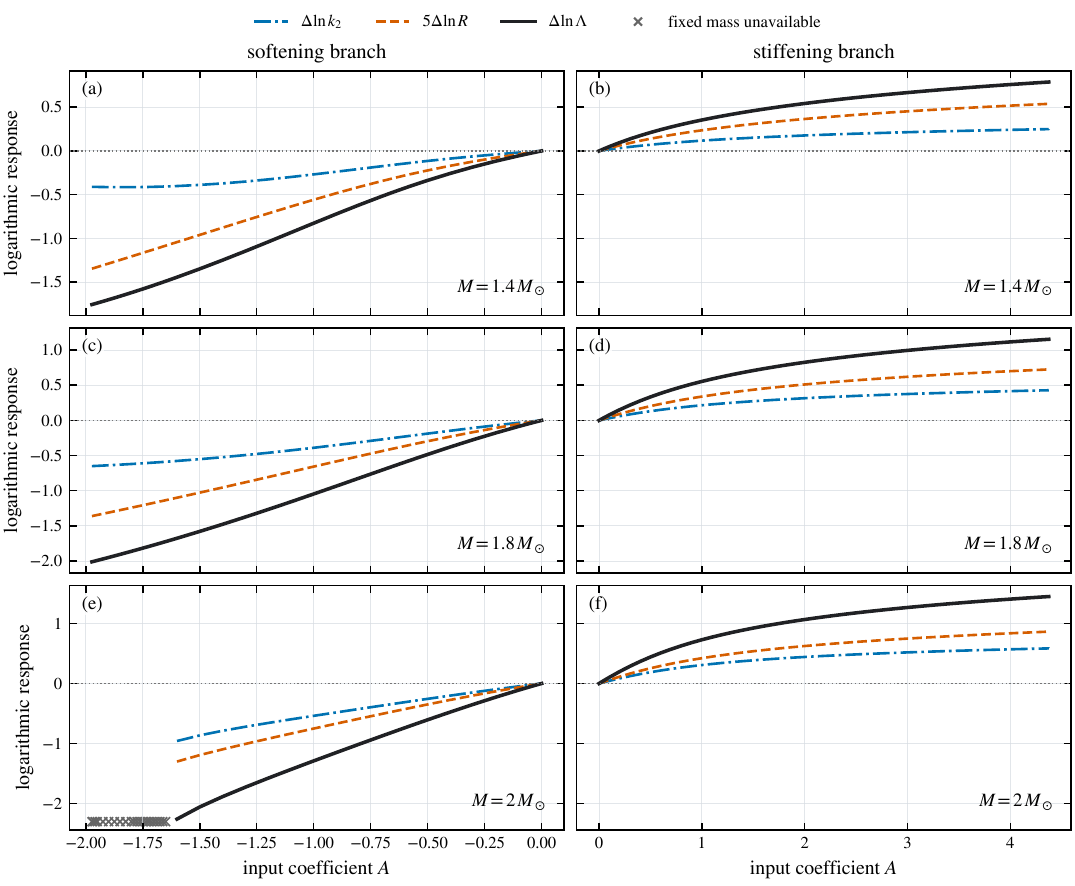}
  \caption{Exact fixed-mass tidal bookkeeping for both signs at \(1.4\),
  \(1.8\), and \(2.0\,M_{\odot}\).  The sum of \(\Delta\ln k_2\) and
  \(5\Delta\ln R\) equals \(\Delta\ln\Lambda\) to the stated residual.
  Crosses retain unavailable \(2.0\,M_{\odot}\) softened rows.  The identity is
  algebraic bookkeeping, not a separate tidal-validation test.}
  \label{fig:tidal-decomposition}
\end{figure*}

At fixed mass, the exact identity
\(\Delta\ln\Lambda=\Delta\ln k_2+5\Delta\ln R\) provides useful
bookkeeping.  Figure~\ref{fig:tidal-decomposition} checks the identity over
both branches with maximum residual \(\TidalBookkeepingResidual\).  This is
not an independent tidal-solver validation and does not separate physical
mechanisms, because \(k_2\) depends on compactness and on the complete stellar
profile.

\section{Numerical robustness, domain reach, and reproducibility}
\label{app:numerical-validation}

The derivative checks compare the continuous reconstructed quantities with
derivatives of independent piecewise cubic Hermite interpolating polynomial
(PCHIP) representations \(P_h(\epsilon)\) and
\(n_{{\rm B},h}(\epsilon)\) \cite{FritschButland1984}.  With
\(\mu_h^{(d)}=(d n_{{\rm B},h}/d\epsilon)^{-1}\), the reported dimensionless
residuals are
\begin{equation}
\begin{aligned}
r_{\rm FL}
  &=\mu_{\rm B}^{\rm eff}\frac{d n_{{\rm B},h}}{d\epsilon}-1,\\
r_{\rm E}
  &=\frac{P-(n_{\rm B}\mu_h^{(d)}-\epsilon)}
  {\max(|P|,|n_{\rm B}\mu_h^{(d)}|,|\epsilon|)},\\
r_c&=c_{\mathrm{s}}^2-\frac{dP_h}{d\epsilon}.
\end{aligned}
\label{eq:numerical-closure-residuals}
\end{equation}
The global summary uses every finite grid node and its maxima are the primary
physical-domain closure result.  The masked smooth-region summary excludes the
first and last four nodes, the complete match-to-ramp interval enlarged by
three upper-grid spacings at either end, and seven-node bands centered on each
of the two retained source-fit transitions.  These masks affect only the
derivative diagnostic: no EoS value is removed or repaired.  The first-law
and Euler-form residuals share the same differentiated density profile and
are algebraically related, whereas \(r_c\) is independent.  These checks apply
the standard cold-fluid identities \cite{ComposeManual2022,OBoyle2020}.

Tables~\ref{tab:production-numerical-settings-thermo}
and~\ref{tab:production-numerical-settings-stellar}, together with
Table~\ref{tab:production-numerical-settings-diagnostics}, record the frozen
settings used by the final campaign.  They are reported explicitly because
internal stage names alone are insufficient for reproduction; exact source
identities remain in the machine-readable provenance records.

\input{tables/production_numerical_settings.tex}

The floating-point physical checks and formal interval are separate layers.
The saved campaign recomputed the formal bounds and required interval
membership before stellar work.  Thus the exact open lower endpoint passed
the floating-point check but was not sent to a stellar calculation.

The stellar evaluation retained a \(c_{\mathrm{s}}^2\geq10^{-10}\) numerical
safety floor, but it was inactive: the smallest saved value over the accepted
campaign was \(3.815\times10^{-5}\).  It therefore did not repair or alter an
accepted EoS.

\input{tables/numerical_validation_manuscript.tex}

\begin{figure*}[!t]
  \centering
  \includegraphics[width=0.78\textwidth]{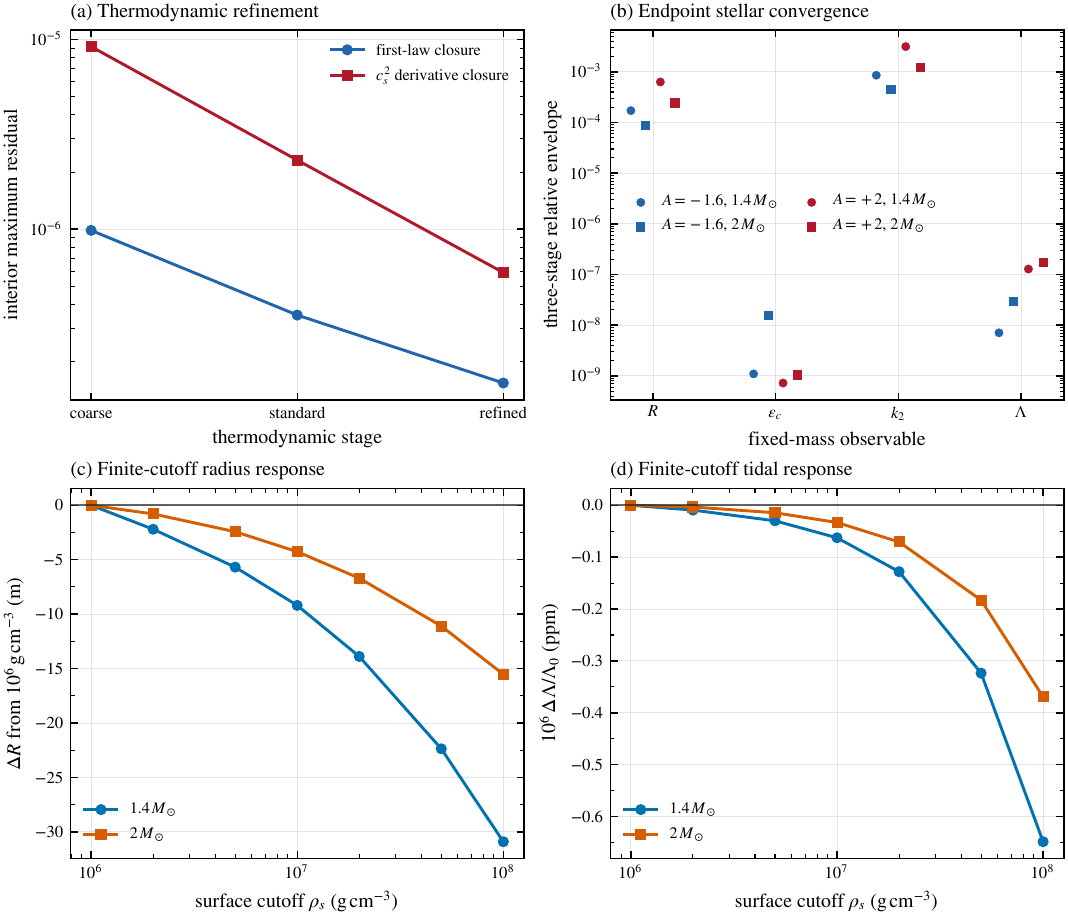}
  \caption{Numerical checks from the retained records.  Panels (a,b) show
  thermodynamic residuals and stellar refinement envelopes; panels (c,d)
  show the direct-baseline response to the supported finite surface cutoff.
  The envelopes are numerical-stage changes, not statistical uncertainties.}
  \label{fig:numerical-validation}
\end{figure*}

Figure~\ref{fig:numerical-validation} combines the thermodynamic and stellar
refinement records with the direct-baseline, in-domain cutoff scan.  As stated
in the main text, that scan does not determine a physical \(P=0\) correction.

\begin{figure*}[!t]
  \centering
  \includegraphics[width=0.90\textwidth]{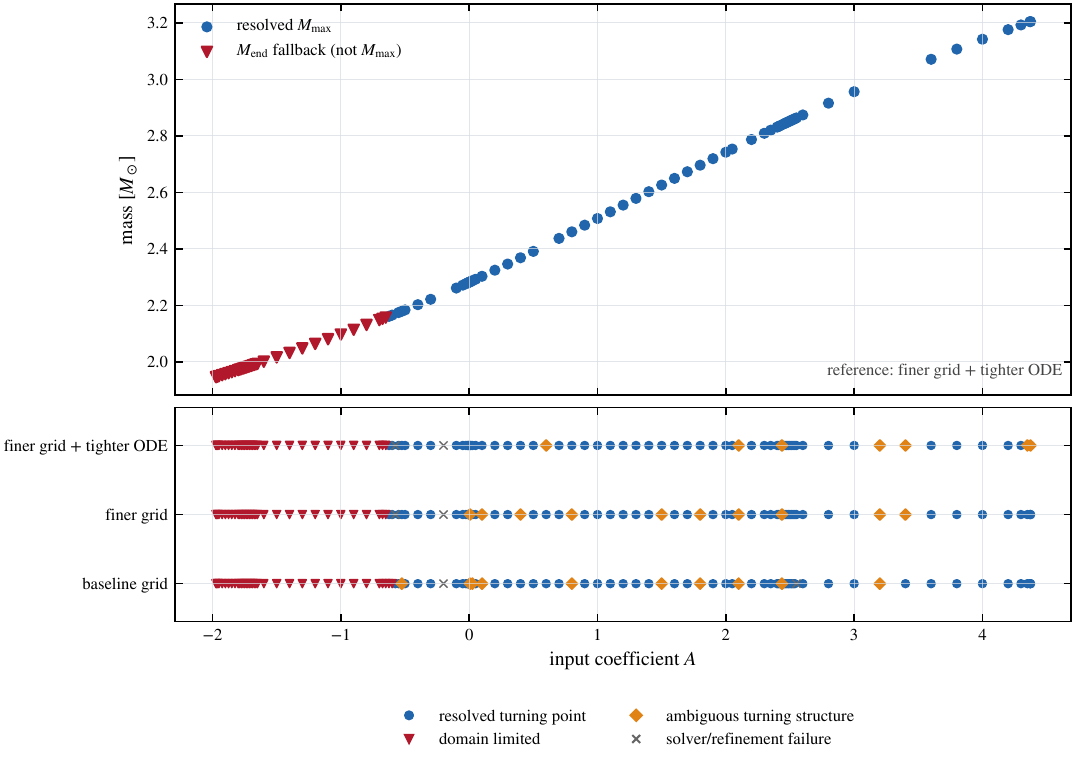}
  \caption{Turning-point and domain status.  The upper panel uses the
  tightest-ODE reference stage: blue circles are refined \(M_{\max}\), while
  red triangles are saved retained-domain masses \(M_{\rm end}\), explicitly
  not \(M_{\max}\).  Ambiguous and failed cases leave gaps.  The lower panel
  shows the status class at all three numerical stages; no continuous trend
  is inferred across gaps.}
  \label{fig:maximum-mass-screening}
\end{figure*}

Figure~\ref{fig:maximum-mass-screening} distinguishes a resolved turning
point from exhaustion of the retained analytic domain and from an
ambiguous or failed refinement.  At the reference tightest-ODE stage,
\(\ReferenceResolvedMaxima\) amplitudes have a refined turning point,
\(\ReferenceDomainLimited\) stop at \(M_{\rm end}\),
\(\ReferenceAmbiguousMaxima\) are ambiguous, and
\(\ReferenceFailedMaxima\)
fail the refinement check.  The last two classes have no mass marker in the
upper panel.  A missing \(M_{\max}\) is therefore unavailable, not evidence
for a lower maximum mass.

The compiled-Ioffe comparison tests implementation of the analytic BSk24
representation \cite{PotekhinIoffe}.  At \(1.4\,M_\odot\), the saved
cutoff-defined \(A=0\) values are \(R=12.602\ \mathrm{km}\),
\(k_2=0.09343\), and \(\Lambda=524.4\).  Published unified-BSk24 comparisons
give \(R=12.59\ \mathrm{km}\) \cite{Perot2019,Perot2020} and independently
\(R=12.547\ \mathrm{km}\), \(\Lambda=522.58\)
\cite{Kunjipurayil2022}; relative to the latter, the saved values differ by
\(0.44\%\) and \(0.35\%\).  This supports the baseline scale but is not a
precision benchmark because the surface and representation conventions are
not matched.  The separate difference between the published rounded
\(R_{1.4}=12.57\ \mathrm{km}\) \cite{Pearson2018} and the
\(12.590\ \mathrm{km}\) CompOSE result \cite{ComposeManual2022} remains an
external provenance or convention question.

All displayed results derive from immutable, checksum-verified records; the
manuscript builders call no scientific solver.  The public ledger records the
identities, source and environment fingerprints, builder inputs, output
hashes, and table provenance
\cite{Papathanasiou2026CampaignData}.  Its archive contains the compact
admission, fixed-mass, radial-support, turning-point, and tidal tables together
with the unprinted finite-step audit and full profiles.  This revision changes
wording, notation, and display precision only; it does not alter the archived
scientific values.  The original raw phase packets remain privately retained
as described in the data and code availability statement.

%% file: tables/production_numerical_settings.tex
\begin{table*}[!t]
\centering
\caption{Frozen production settings for the thermodynamic and formal-bound
stages of the final campaign.  Node counts on either side of the match include
the shared match node.  Tolerances are absolute unless explicitly labeled
relative.  Boundary values are rounded for display; the archived configuration
retains the full numerical literals.}
\label{tab:production-numerical-settings-thermo}
\renewcommand{\arraystretch}{1.03}
\small
\begin{tabular}{@{}p{0.24\textwidth}p{0.70\textwidth}@{}}
\toprule
Layer & Frozen campaign setting \\
\midrule
Thermodynamic grids &
Coarse, standard, and refined grids use
\((N_{<},N_{>})=(1025,2049),(2049,4097),(4097,8193)\), giving
3073, 6145, and 12289 total nodes.  Below
\(\epsilon_{\rm match}\), nodes are geometric from
\(5.6096\times10^{-7}\) to
\(80\ \mathrm{MeV\,fm^{-3}}\); above it they are linear through
\(1.509\times10^3\ \mathrm{MeV\,fm^{-3}}\). \\
Raw full-domain gate &
The dense gate uses \((4097,16385)\), or 20481 total nodes.  Candidate
extrema are refined by bounded scalar minimization with
\(x_{\rm atol}=10^{-11}\); sampled \(c_{\mathrm{s}}^2=0\) and \(1\) crossings use
Brent roots with \(x_{\rm tol}=r_{\rm tol}=10^{-12}\). \\
{\raggedright Formal amplitude bounds\par} &
The baseline reconstruction uses \((4097,8193)\).  A separate continuous
baseline precheck uses \((16385,32769)\), or 49153 total nodes.  The affected
domain begins with 32769 search nodes and contains 32771 after mandatory-point
insertion; log-ratio minima use bounded
\(x_{\rm atol}=10^{-10}\), followed by a
\(5\times10^{-12}\) verification tolerance. \\
\bottomrule
\end{tabular}
\end{table*}

\begin{table*}[!t]
\centering
\caption{Frozen production settings for the stellar and turning-point stages of
the final campaign.  Here \(N_r\) is the number of saved radial rows.
Tolerances are absolute unless explicitly labeled relative.}
\label{tab:production-numerical-settings-stellar}
\renewcommand{\arraystretch}{1.03}
\small
\begin{tabular}{@{}p{0.24\textwidth}p{0.70\textwidth}@{}}
\toprule
Layer & Frozen campaign setting \\
\midrule
Stellar stages &
Standard, finer-grid, and tighter-ODE-tolerance stages use
\((N_P,r_{\rm tol},a_{\rm tol},N_r)=(61,10^{-8},10^{-10},601)\),
\((121,10^{-8},10^{-10},601)\), and
\((121,10^{-10},10^{-12},1201)\).  Central pressures are geometric from
\(2\ \mathrm{MeV\,fm^{-3}}\) to the retained endpoint pressure.  The
Runge--Kutta integration uses \(r_{\min}=10^{-4}\ \mathrm{km}\) and
\(r_{\max}=25\ \mathrm{km}\).  Successful rows additionally require
\(M\geq0.05M_\odot\), \(R\geq3\ \mathrm{km}\), and compactness below
\(4/9\). \\
Fixed-mass roots &
The first adjacent true bracket on the successful pre-peak sequence is solved
in central pressure with Brent's method,
\(x_{\rm tol}=10^{-7}\ \mathrm{MeV\,fm^{-3}}\) and relative tolerance
\(4\epsilon_{\rm mach}\simeq8.882\times10^{-16}\).  Interior root evaluations omit tides;
one final tidal/profile calculation is made at the root. \\
{\raggedright Turning-point refinement\par} &
Cases reaching the recorded \(1.95M_\odot\) screen are assessed for
refinement.  Refinement requires no sampled solver gap and a unique
positive-to-negative secant transition in \(dM/dP_c\).  A 17-point geometric
bracket is followed by bounded optimization in \(\ln P_c\), with
\(x_{\rm atol}=5\times10^{-4}\) and at most 32 iterations.  The refined point
must lie strictly inside the bracket with the required secant signs. \\
\bottomrule
\end{tabular}
\end{table*}

\begin{table*}[!t]
\centering
\caption{Frozen production settings for the stellar surface and diagnostic
layers of the final campaign.  Tolerances are absolute unless explicitly
labeled relative.}
\label{tab:production-numerical-settings-diagnostics}
\renewcommand{\arraystretch}{1.03}
\small
\begin{tabular}{@{}p{0.24\textwidth}p{0.70\textwidth}@{}}
\toprule
Layer & Frozen campaign setting \\
\midrule
{\raggedright Finite surface/tides\par} &
The event is the retained lower-fit boundary,
\(\rho=10^6\ \mathrm{g\,cm^{-3}}\),
\(\epsilon=5.6096\times10^{-7}\) and
\(P=1.354\times10^{-11}\ \mathrm{MeV\,fm^{-3}}\), not
\(P=0\).  The tidal variable starts from \(y=2\) and is integrated separately
over the saved background with the same ODE tolerances.  Its terminal value at
the cutoff is supplied directly to the exterior Love-number formula under the
continuous-hadronic convention.  Required jumps must be applied exactly once;
the solver capability record declares no physical surface-density jump and no
retained discontinuities, so no finite-density surface-jump correction is
applied.  This declaration selects the bookkeeping convention and does not
imply \(\epsilon(R)=0\) at the retained cutoff. \\
Derivative masks &
Closure residuals use the masks stated below.  Advisory derivatives use the
core \(\epsilon\geq\epsilon_{\rm match}\), omit the first and last two
profile nodes and the match node plus two neighbors on each side, and use
second-order logarithmic finite differences for the fundamental derivatives.
The \(\Gamma_{\rm eq}<4/3\) test subtracts the scale-local allowance
\(64\epsilon_{\rm float}\max(1,\max|\Gamma_{\rm eq}|)\); the two fundamental-
derivative sign tests use strict comparison with zero.  These diagnostics
have no acceptance authority. \\
\bottomrule
\end{tabular}
\end{table*}

%% file: tables/numerical_validation_manuscript.tex
\begin{table*}[t]
\centering
\caption{Compact numerical checks.  Stellar envelopes are relative ranges
across three saved numerical stages, not statistical uncertainties.}
\label{tab:numerical-validation}
\renewcommand{\arraystretch}{0.96}
\footnotesize
\begin{tabular}{p{0.27\textwidth}p{0.66\textwidth}}
\toprule
Check & Saved result \\
\midrule
{\raggedright Ramped-Gaussian pressure primitive\par} & Maximum \(|P_{\rm analytic}-P_{\rm quad}|\) at 11 nodes per endpoint: \(4.26\times 10^{-14}\,\mathrm{MeV\,fm^{-3}}\), compared with the \(5.00\times 10^{-12}\,\mathrm{MeV\,fm^{-3}}\) allowance. \\
BSk24 benchmark & Independent compiled Ioffe reference: \(|\Delta R_{1.4}|=1.08\times 10^{-4}\,\mathrm{km}\); maximum pre-peak mass--radius curve separation over \(1.0\)--\(2.2\,M_\odot\): \(3.35\times 10^{-4}\,\mathrm{km}\).  These submeter separations are consistent with the documented few-ppm coefficient-literal difference between the official routine and the exact-decimal transcription, together with interpolation along the independently sampled mass--radius sequences. \\
Zero-deformation control & Direct BSk24 and \(A=0\) are array-exact for the retained thermodynamic fields and saved \(M\), \(R\), and \(\Lambda\) sequences; maximum residual: zero. \\
Thermodynamic closure & Masked smooth-region maxima (first law, Euler form, \(c_{\mathrm{s}}^2\) derivative): \(( 1.55\times 10^{-7},\,1.55\times 10^{-7},\,5.91\times 10^{-7} )\); the primary full-domain maxima are \(( 3.48\times 10^{-6},\,3.48\times 10^{-6},\,1.87\times 10^{-5} )\).  The first two are algebraically related; the third is distinct. \\
Fixed-mass roots & Maximum \(|M-M_{\rm target}|\) at the tightest ODE tolerances, for both endpoints and \(1.4,2.0\,M_\odot\): \(3.63\times 10^{-12}\,M_\odot\). \\
Endpoint stellar refinement & Maximum relative envelopes \((R,\epsilon_c,k_2,\Lambda)\): \(( 6.32\times 10^{-4},\,1.55\times 10^{-8},\,3.15\times 10^{-3},\,1.70\times 10^{-7} )\). \\
Near-domain stress test & At \(A=-1.6,\ M=2.0\,M_\odot\), relative envelopes \((\epsilon_c,\Lambda)\): \(( 1.55\times 10^{-8},\,2.97\times 10^{-8} )\). \\
Finite surface cutoff & Raising the one-sided cutoff from \(10^6\) to \(10^8\,\mathrm{g\,cm^{-3}}\): maximum \(|\Delta R|=0.0309\,\mathrm{km}\), \(|\Delta k_2|=1.154\times10^{-3}\) (\(1.235\%\)), and \(|\Delta\Lambda/\Lambda_0|=0.648\) ppm.  The percent-level \(\Delta\ln k_2\) and \(5\Delta\ln R\) terms nearly cancel, leaving the much smaller relative tidal response.  No \(P=0\) extrapolation. \\
\bottomrule
\end{tabular}
\end{table*}